\documentclass[smallextended]{svjour3}
\usepackage[dvips]{graphicx}
\usepackage{epsfig,amssymb,amsmath,color}
\newcommand{\beq}{\begin{equation}}
\newcommand{\eeq}{\end{equation}}
\newcommand{\beqn}{\begin{eqnarray}}
\newcommand{\eeqn}{\end{eqnarray}}

\def\bmath#1{\mbox{\boldmath$#1$}}

\long\def\symbolfootnote[#1]#2{\begingroup%
\def\thefootnote{\fnsymbol{footnote}}\footnote[#1]{#2}\endgroup}
\usepackage{txfonts}
\usepackage{natbib}
\bibpunct{(}{)}{;}{a}{}{,}

\journalname{Exp. Astro.}
\begin{document}
\title{Estimation of Radio Interferometer Beam Shapes Using Riemannian Optimization}
   \author{Sarod Yatawatta
          }

   \institute{ ASTRON, Dwingeloo, NL\\
          \email{yatawatta@astron.nl}
}
\date{Draft version. The final publication is available at springerlink.com}
\maketitle

\begin{abstract}
The knowledge of receiver beam shapes is essential for accurate radio interferometric imaging. Traditionally, this information is obtained by holographic techniques or by numerical simulation. However, such methods are not feasible for an observation with time varying beams, such as the beams produced by a phased array radio interferometer. We propose the use of the observed data itself for the estimation of the beam shapes. We use the directional gains obtained along multiple sources across the sky for the construction of a time varying beam model. The construction of this model is an ill posed non linear optimization problem. Therefore, we propose to use Riemannian optimization, where we consider the constraints imposed as a manifold. We compare the performance of the proposed approach with traditional unconstrained optimization and give results to show the superiority of the proposed approach.
\end{abstract}
%\begin{keywords}
%
%Radio Interferometry, Manifold Optimization, Array Processing.
%\end{keywords}

\section{Introduction}
Most interferometric observations are done using receivers that are more sensitive towards a part of the sky. This narrow field of view is attained using directive antennas (such as a dish) or by beamforming. Due to this reason, images made by such interferometric observations are distorted, with the distortion increasing for celestial objects further away from the direction where the beams are pointed at. Therefore, the knowledge of the beam shape is essential to correct for this distortion while producing accurate and distortion free images. Traditionally beam information is obtained  by  holographic techniques \citep{Scott,Bennet,Popping}  or by drift scanning \citep{PAPER}.  These methods work well for a stationary and stable beam pattern, such as the beams produced by movable dish based receivers. However, such techniques will not give accurate results for interferometers that have time varying beam shapes. A case in point is the beam shapes produced by phased array radio telescopes such as LOFAR\footnote{The Low Frequency Array: http://www.lofar.org}. During an observation with a phased array, the beamforming weights change and this results in a variation of the overall beam pattern. In addition, different  element layouts between different stations also make the beam shape significantly different. Moreover, secondary effects such as mutual coupling would make the beam shapes different for each receiver.

In this paper, we propose to use the observation itself to extract beam shape information rather than using a priori information such as by holography or drift scanning. Efficient techniques are available to extract directional gains along multiple directions in the sky \citep{SAGE,Kaz2}. We can extract these gains not only along the direction where the beams are pointed at, but also along other directions where there are well known celestial sources to sufficiently sample the beam shape. Once these directional gains are available, the recovery of the beam shape is an ill posed nonlinear optimization problem. Therefore, we have to apply additional constraints to get a satisfactory solution. This naturally leads us to optimization on a Riemannian manifold, as discussed in \cite{Gabay,manton2004}. As a byproduct of this process, we can also obtain intrinsic fluxes of the celestial sources used in calibration, subject to provision of a few known sources for absolute flux calibration. 

Manifold optimization has been applied in diverse areas of research and a complete overview is given in \cite{AMS}. In this paper, we present a hybrid optimization method that jointly uses steepest descent (SD) and the Broyden Fletcher Goldfarb Shanno (BFGS) algorithms on a Riemannian manifold.  We use the geodesic stepping method \citep{Fiori} based on a Riemannian gradient as our Riemannian steepest descent (RSD) method. However, SD method has the drawback of only having linear convergence rate, especially close to the solution. To accelerate the convergence, we use the Riemannian BFGS \citep{RBFGS}  algorithm in conjunction  with the RSD method. 

Together with calibration along multiple directions \citep{SAGE,Kaz2}, the methods proposed in this paper to estimate the beam shape and intrinsic fluxes can also be considered as one cycle of self-calibration, where we also update the sky model. However, in this paper we focus our attention on the latter part of the self-calibration cycle, i.e., the estimation of beam shapes and the estimation of intrinsic fluxes.

The rest of the paper is organized as follows: In section \ref{interferometry}, we give an overview of radio interferometry. Next in section \ref{beamest}, we present the beam shape estimation approach and we elaborate on Riemannian optimization in section \ref{rmopt}. In section \ref{simul}, we give simulation results to verify the proposed approach and give conclusions in section \ref{conc}.

Notation: Matrices and vectors are denoted by bold upper and lower case letters as ${\bf B}$ and ${\bf v}$, respectively. The canonical vector with a one at the $p$-th location and zeros everywhere else is given by ${\bf e}_p$. The transpose, Hermitian transpose and conjugation are given by $(.)^T$, $(.)^H$ and $(.)^{\star}$, respectively. The matrix Kronecker product is denoted by $\otimes$ and the Frobenius norm is given by $\|.\|$. The set of complex numbers is denoted by ${\mathbb C}$.

\section{Radio Interferometry\label{interferometry}}
We give a brief overview of radio interferometry and calibration in this section. Consider an interferometer formed by station $p$ and station $q$. The received data after correlation and correction for delay errors between stations can be given as
\beq\label{vis}
{\bf V}_{pq}=\sum_{m=1}^{M} {\bf J}_{pm} \widetilde{\bf C}_{pqm}{\bf J}_{qm}^H +{\bf N}_{pq}
\eeq
where ${\bf V}_{pq}$ ($\in {\mathbb C}^{2\times 2}$) is the visibility matrix \citep{HBS} and ${\bf N}_{pq}$ ($\in {\mathbb C}^{2\times 2}$) is the noise. In (\ref{vis}), the observation consists of radiation from $M$ discrete sources in the sky, whose coherencies are given by $\widetilde{\bf C}_{pqm}$ ($\in {\mathbb C}^{2\times 2}$). For a point source with intensity $I_m$ and polarized flux $Q_m,U_m,V_m$, the coherency for linearly polarized receptors is given by
\beq\label{coh}
\widetilde{\bf C}_{pqm}=e^{j\phi_{pqm}}\left[ \begin{array}{cc}
I_m+Q_m & U_m+jV_m\\
U_m-jV_m & I_m-Q_m
\end{array} \right]
\eeq
where $\phi_{pqm}$ is the Fourier phase component that depends on the direction in the sky as well as the separation of station $p$ and $q$. In calibration, we estimate the Jones matrices ${\bf J}_{pm}$ ($\in {\mathbb C}^{2\times 2}$) for each station as well as for each direction in the sky \citep{SAGE,Kaz2}. For each source, we have accurate knowledge of the directions (or positions) in the sky and only an apparent knowledge of the fluxes. The solutions obtained for ${\bf J}_{pm}$ contain the information about the beam shape along each direction. However, as noted in \cite{H4}, there is always an ambiguity in these solutions and therefore, we cannot use the values of ${\bf J}_{pm}$ to directly construct a beam model. Most celestial sources are unpolarized and thus, there is a unitary ambiguity in the solutions and what we obtain is ${\bf J}_{pm}{\bf U}_m$ where ${\bf U}_m$ ($\in {\mathbb C}^{2\times 2}$) is an unknown unitary matrix.

\section{Beam Shape Estimation\label{beamest}}
The beam shape of a phased array receiver consists of two parts. Each element used in beamforming has the {\em element} beam pattern that is sensitive to the full sky. Using beamforming, this beam is narrowed down to cover the field of interest in the sky. Therefore, for the $p$-th station, the beam gain along the $m$-th direction can be given as $\gamma_{pm}{\bf E}_{pm}$ where the element beam is given by ${\bf E}_{pm}$  ($\in {\mathbb C}^{2\times 2}$). What we are interested in is the array gain $\gamma_{pm}$ ($\in {\mathbb C}$), which is dependent on the beamforming weights and is changing as the weights change. This is $1$ at the direction where the beam is pointed at and as the sky rotates, due to the continuous tracking of a single direction in the sky, the overall values for $\gamma_{pm}$ change with time.

Note that $\gamma_{pm}$ is a complex valued parameter: Therefore, in general we estimate the voltage beam shape. Moreover, any atmospheric phase variations are also incorporated into the value of $\gamma_{pm}$. Before we proceed, we make the following assumptions:
\begin{itemize}
\item We have satisfactory knowledge of the element beam pattern ${\bf E}_{pm}$, mainly by numerical simulation.
\item Although the calibration of (\ref{vis}) was performed using apparent fluxes of the $M$ sources, we assume approximate knowledge of intrinsic fluxes of at least a few sources (we call them as {\em seed} sources).
\item We assume perfect knowledge of source positions and assume atmospheric phase errors are absorbed  into the solutions of ${\bf J}_{pm}$, in addition to beam shape errors.
\end{itemize}

We select a set of complex valued basis functions to model the beam shape.  Let the number of basis functions be $D$ and then we can evaluate the beam gain as
\beq\label{beamval}
 {\gamma}_{pm}={\bf e}_p^T{\bf B}{\bf b}_m
\eeq
where ${\bf B}$ ($\in {\mathbb C}^{N\times D}$) gives the beam model for all $N$ stations. The values of the basis functions along the $m$-th direction is given by ${\bf b}_m$ ($\in {\mathbb C}^{D\times 1}$). The canonical vector is given as 
\beq
{\bf e}_p\buildrel\triangle\over=[0,0,\ldots,0,1,0,\ldots,0]^T
\eeq
with all zeros except a $1$ at the $p$-th location.
Ideally, for a source $m$ with perfect knowledge of its fluxes we get 
\beq \label{ideal}
{\bf C}_{pqm}\gamma_{pm}\gamma_{qm}^{\star}={\bf J}_{pm} \widetilde{\bf C}_{pqm}{\bf J}_{qm}^H
\eeq
where ${\bf C}_{pqm}$ ($\in {\mathbb C}^{2\times 2}$) is the true coherency, taking into account the element beam shapes ${\bf E}_{pm}$ and ${\bf E}_{qm}$. Note that the ambiguities in ${\bf J}_{pm}$ and ${\bf J}_{qm}$ cancel out and does not affect (\ref{ideal}). 

Using (\ref{beamval}) and (\ref{ideal}), the  cost function that needs to be minimized to estimate ${\bf B}$ can be given as
\beq\label{fcost}
f({\bf B})=\sum_{p,q,m} \|{\bf C}_{pqm}\gamma_{pm}\gamma_{qm}^{\star}-{\bf J}_{pm} \widetilde{\bf C}_{pqm}{\bf J}_{qm}^H\|^2
\eeq
where the summation is taken over all baselines $p$,$q$ and all sources $m$ whose intrinsic fluxes are known.  The minimization of (\ref{fcost})  to yield an estimate for ${\bf B}$ is highly ill posed mainly due to not having enough sources with known fluxes as well as sufficient intensities to yield a good solution for ${\bf J}_{pm}$ under noisy observations. Therefore, we impose the additional constraint that preserves the total power received by all stations. As shown in appendix \ref{proof_alpha},  the total power constraint can be represented as
\beq \label{alpha}
trace({\bf B}^H{\bf B}) =\alpha
\eeq
where $\alpha$ is a fixed real value. Although we cannot exactly determine $\alpha$, a nominal value based on the chosen basis functions and a nominal beam shape is sufficient.

Apart from the cost function (\ref{fcost}), we need the gradient of the cost function in our optimization routines. Using techniques of \citep{Mdiff}, we get the derivative as (proof is given in appendix \ref{proof_gradf})
\beq\label{gradf}
\frac{\partial f}{\partial {\bf B}}=\sum_{p,q,m} \beta_{1pqm} {\bmath \Gamma}_{1pqm}-\beta_{2pqm} {\bmath \Gamma}_{2pqm}-\beta_{3pqm} {\bmath \Gamma}_{3pqm}
\eeq
where
\beqn \label{gradfterms}
\beta_{1pqm}&\buildrel\triangle\over=&trace\left({\bf C}_{pqm}^H{\bf C}_{pqm}\right),\\\nonumber
\beta_{2pqm}& \buildrel\triangle\over=&trace\left({\bf C}_{pqm}^H{\bf J}_{pm} \widetilde{\bf C}_{pqm}{\bf J}_{qm}^H\right),\\\nonumber
\beta_{3pqm}& \buildrel\triangle\over=&trace\left(({\bf J}_{pm} \widetilde{\bf C}_{pqm}{\bf J}_{qm}^H)^H{\bf C}_{pqm}\right),\\\nonumber
{\bmath \Gamma}_{1pqm} & \buildrel\triangle\over=& {\bf e}_q{\bf b}_m^T({\bf b}_m^H{\bf B}^H{\bf e}_p)({\bf e}_p^T{\bf B}{\bf b}_m)({\bf b}_m^H{\bf B}^H{\bf e}_q)\\\nonumber
&&\mbox{}+{\bf e}_p{\bf b}_m^T({\bf b}_m^H{\bf B}^H{\bf e}_p)({\bf e}_q^T{\bf B}{\bf b}_m)({\bf b}_m^H{\bf B}^H{\bf e}_q),\\\nonumber
{\bmath \Gamma}_{2pqm}& \buildrel\triangle\over=&{\bf e}_q{\bf b}_m^T({\bf b}_m^H{\bf B}^H{\bf e}_p),\\\nonumber
{\bmath \Gamma}_{3pqm}& \buildrel\triangle\over= &{\bf e}_p{\bf b}_m^T({\bf b}_m^H{\bf B}^H{\bf e}_q).
\eeqn

To summarize: we need to estimate ${\bf B}$ by minimizing the cost function (\ref{fcost}) subject to the constraint (\ref{alpha}). As described in next section, we choose Riemannian optimization to solve this problem.
\section{Riemannian Optimization\label{rmopt}}
We give a brief description of the motivation behind using Riemannian optimization as opposed to traditional constrained optimization. As presented in \cite{Gabay,AMS} and other work, traditional constrained optimization (using Lagrange multipliers) increases the dimensionality of the problem, therefore making it more complicated. On the other hand, the constraints ((\ref{alpha}) in our case) can be thought of as restricting ${\bf B}$ onto a Riemannian manifold. Therefore the dimensionality is not increased. However, the traditional gradient based optimization algorithms applicable in Euclidean space cannot be applied in a straight forward manner because the tangent planes change depending on the value of ${\bf B}$.

We present two optimization algorithms on the manifold (\ref{alpha}) that we use in beam estimation. The first one is RSD, as presented in \citep{Fiori}  and the second one is RBFGS, as presented in \citep{RBFGS}. The steepest descent method has linear convergence but simpler to implement while the RBFGS has super-linear convergence. Apart from the cost function (\ref{fcost}), the only requirement is the gradient (\ref{gradf}). By using both algorithms in a hybrid fashion, we have faster convergence and are less susceptible to get stuck in a local minimum.

\subsection{Riemannian Steepest Descent Algorithm}
We choose the method proposed in \cite{Fiori} as our RSD algorithm.  
We briefly give the algorithm that we use for estimating ${\bf B}$, more detail can be found in \cite{Fiori}.
\begin{enumerate}
\item Calculate $\frac{\partial f}{\partial {\bf B}}$ using (\ref{gradf}) and (\ref{gradfterms}).
\item Calculate the Riemannian gradient 
\beq\label{rgrad}
-\nabla_{\bf B} f=\frac{\partial f}{\partial {\bf B}}-\frac{1}{\alpha}{\bf B} real\left(trace({\bf B}^H\frac{\partial f}{\partial {\bf B}})\right).
\eeq
\item Find the step size $h$ in $[0,2\pi/\omega]$, $\omega=\|\nabla_{\bf B} f\|/\sqrt{\alpha}$ that minimizes
\beq
f\left({\bf B}(h)\right)=f\left({\bf B}\cos(\omega h)+(\nabla_{\bf B}f) \sin(\omega h)/\omega\right).
\eeq
\item Update ${\bf B}\gets {\bf B}\cos(\omega h)+(\nabla_{\bf B}f) \sin(\omega h)/\omega$.
\item If $\|\nabla_{\bf B} f\|$ is too small or the maximum number of iterations has reached stop, else go back to step 1.
\end{enumerate}

\subsection{Riemannian Broyden Fletcher Goldfarb Shanno Algorithm}
In order to present the RBFGS algorithm, we use an alternative representation for the beam model ${\bf B}$ as follows 
\beq
{\bf x}\buildrel\triangle\over = [vec(real({\bf B}))^T vec(imag({\bf B}))^T]^T/\sqrt{\alpha}
\eeq
where ${\bf x}$ is a real vector of size $2ND\times 1$.
Then, the constraint (\ref{alpha}) can be rewritten as
\beq \label{stf}
{\bf x}^T {\bf x}=1
\eeq
which makes ${\bf x}$ restricted to a (real) Stiefel manifold of size $2ND\times 1$ and dimension $2ND-1$. This also means that ${\bf x}$ is on a $2ND-1$ dimensional unit sphere (which is a special case of a Stiefel manifold). 
We adopt the BFGS algorithm on a unit sphere as presented in \cite{RBFGS}. In order to fully implement this algorithm, we need to define several operators on the manifold.
The projection of any vector ${\bmath \eta}$ to tangent space at ${\bf x}$ on the manifold is given by
\beq \label{r_proj}
{\bf P}_{\bf x}({\bmath \eta})\buildrel\triangle \over =({\bf I}-{\bf x}{\bf x}^T){\bmath \eta}.
\eeq

The cost function (\ref{fcost}) can be expressed as $f({\bf x})=f({\bf B})$ with some abuse of notation. The gradient is constructed from (\ref{gradf}) by projecting it onto the tangent space as
\beq
grad(f({\bf x}))=({\bf I}-{\bf x}{\bf x}^T)[vec(real(\frac{\partial f}{\partial {\bf B}}))^T vec(imag( \frac{\partial f}{\partial {\bf B}}))^T]^T/\sqrt{\alpha}.
\eeq

The retraction of vector ${\bmath \eta}$ in the tangent space at ${\bf x}$ to the manifold is given by
\beq
{\bf R}_{\bf x}({\bmath \eta}) \buildrel\triangle\over = \frac{({\bf x}+{\bmath \eta})}{\|({\bf x}+{\bmath \eta})\|}.
\eeq

In  \cite{RBFGS}, vector transport is used to transport a tangent vector from a tangent space at one point to the tangent space at another point on the manifold. This operator is given by
\beq
{\bf T}_{\bf x}({\bmath \eta},{\bmath \zeta})\buildrel \triangle\over =\left({\bf I}-\frac{({\bf x}+{\bmath \eta})({\bf x}+{\bmath \eta})^T}{\| ({\bf x}+{\bmath \eta}) \|^2}\right) {\bmath \zeta}
\eeq
and its inverse (inverse vector transport) is given by
\beq
{\bf T}_{\bf x}^{-1}({\bmath \eta},{\bmath \zeta})\buildrel\triangle\over = \left({\bf I}-\frac{({\bf x}+{\bmath \eta}) {\bf x}^T}{{\bf x}^T ({\bf x}+{\bmath \eta})} \right) {\bmath \zeta}.
\eeq

With these definitions at hand, we are ready to implement the RBFGS algorithm.
\begin{itemize}
\item Initial conditions: Hessian approximation ${\bf H}_1={\bf I}$.
\item Iterations $k=1$ to $max\ iterations$ 
\begin{enumerate}
\item Obtain ${\bmath \eta}_k$ by solving ${\bf H}_k {\bmath \eta}_k=-grad f({\bf x}_k)$.
\item Perform line search: set $a=1$; $c=grad f({\bf x}_k)^T {\bmath \eta}_k$
\begin{itemize}
\item while  $f({\bf R}_{{\bf x}_k}(2 a {\bmath \eta}_k))-f({\bf x}_k) < a c$,
  update $a \leftarrow 2 a$.
\item while $f({\bf R}_{{\bf x}_k}( a {\bmath \eta}_k))-f({\bf x}_k) >0.5 a c$,
update $a\leftarrow 0.5 a$.
\end{itemize}
\item Update ${\bf x}_{k+1} \leftarrow {\bf R}_{{\bf x}_k}(a {\bmath \eta}_k)$.
\item ${\bf s}_k={\bf T}_{{\bf x}_k}(a {\bmath \eta}_k,a {\bmath \eta}_k)$;
${\bf y}_k=grad f({\bf x}_{k+1})-{\bf T}_{{\bf x}_k}(a {\bmath \eta}_k,grad f({\bf x}_{k}))$
\item Update Hessian approximation as $\widetilde{\bf H}_k={\bf T}({\bf x}_{k+1},a {\bmath \eta}_k) {\bf H}_k {\bf T}^{-1}({\bf x}_{k+1},a {\bmath \eta}_k)$

and ${\bf H}_{k+1}=\widetilde{\bf H}_k-\frac{\widetilde{\bf H}_k {\bf s}_k {\bf s}_k^T \widetilde{\bf H}_k}{{\bf s}_k^T \widetilde{\bf H}_k  {\bf s}_k} +\frac{{\bf y}_k{\bf y}_k^T}{{\bf y}_k^T {\bf s}_k}$.
\end{enumerate}
\end{itemize}

\subsection{Hybrid Optimization}
With the RSD and RBFGS algorithms as implemented above, the implementation of the hybrid algorithm is as follows. 
\begin{enumerate}
\item Start with nominal beam shape ${\bf B}_0$ and $\alpha=trace({\bf B}_0^H {\bf B}_0)$.
\item In parallel, run RSD and RBFGS with maximum number of iterations fixed to  $n_1$ (about $10$). 
\item Compare the final cost from both RSD and RBFGS algorithms. Select the solution with the lowest cost from either RSD or RBFGS as the updated value for ${\bf B}$.
\item If maximum number of hybrid iterations $n_2$ (about $200$) is reached, stop. Else  go back to step 2 with the updated ${\bf B}$ as the initial value.
\end{enumerate}
Note that in this algorithm, we use two limits for the number of iterations, the first one for each RSD and RBFGS iteration limit ($n_1$) and the second one for the hybrid iteration limit ($n_2$). It should also be mentioned that the solution obtained for ${\bf B}$ always has an unknown complex scalar ambiguity. This can be eliminated by normalizing the peak of all the estimated beams to a pure real value.

The initial selection of $\alpha$ is done by assuming a nominal beam model. Depending on additional information such as the beamformed element layout and the frequency of observation, and also depending on the basis functions chosen, it is possible to determine an accurate value for $\alpha$. We also use the nominal beam model as our initial value in optimization.

\subsection{Flux Estimation}
 Once we have the estimate for ${\bf B}$, it is also possible to estimate the intrinsic fluxes for all sources in our calibration model (\ref{vis}).  In order to do this, we make an additional assumption:
\begin{itemize}
\item All stations $p$ see the same intrinsic sky, therefore, for a sky consisting of point sources ${\bf C}_{pqm}={\bf C}_m$ and the common Fourier phase term in (\ref{coh}) and (\ref{ideal}) can be precomputed. For an array with parallel dipoles (such as LOFAR), we assume the element beam pattern of each station is identical. Therefore, the dependence of $p$ and $q$ on ${\bf C}_{pqm}$ is eliminated.
\end{itemize}

Under this assumption, for the $m$-th source we define the cost to be minimized  in order to estimate the flux as
\beq
g_{m}({\bf C}_m)=\sum_{p,q} \|{\bf C}_{m}\gamma_{pm}\gamma_{qm}^{\star}-{\bf J}_{pm} \widetilde{\bf C}_{pqm}{\bf J}_{qm}^H\|^2.
\eeq

By making $\frac{\partial g_{m}({\bf C}_m)}{\partial {\bf C}_m}={\bf 0}$, we get the estimate 
\beq \label{cohest}
\widehat{{\bf C}}_{m}=\frac{\sum_{p,q} \gamma_{pm}^\star\gamma_{qm}{\bf J}_{pm} \widetilde{\bf C}_{pqm}{\bf J}_{qm}^H}{\sum_{p,q} |\gamma_{pm}|^2|\gamma_{qm}|^2}.
\eeq

It should be reminded that we can use (\ref{cohest}) to estimate fluxes for any point source along the direction of which we have obtained a calibration solution. Of course, for sources that are far away from the center of the beam, the denominator of (\ref{cohest}) would get close to zero, making our flux estimate unreliable. This can be overcome by combining observations taken at different epochs.
Once we have updated the sky model using  (\ref{cohest}), we can go back to update our estimate of ${\bf B}$. Therefore, with an updated sky model, we can use more {\em seed} sources to better constrain the estimation of the beam shape. In addition, this step also completes one self-calibration loop.

\section{Simulation Results\label{simul}}
\begin{figure}[htbp]
\begin{minipage}{0.98\linewidth}
\centering
 \centerline{\epsfig{figure=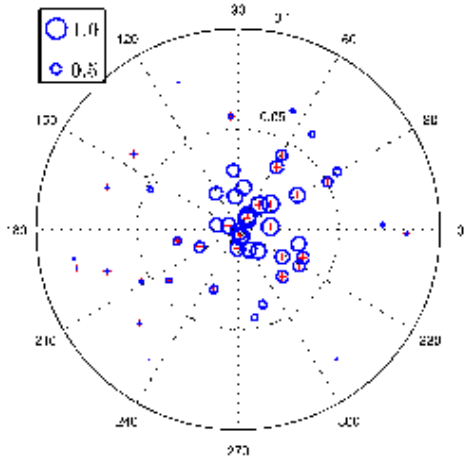,width=10.0cm}}
\caption{Sky model in a field of view of $8$ degrees in diameter. The circles correspond to the ratio between the apparent flux used in calibration and the intrinsic flux of each source.\label{skymod}}
\end{minipage}
\end{figure}

We consider an observation with a field of view of $8$ degrees (diameter) in  the sky. We simulate $M=50$ sources, randomly placed in the field of view with no intrinsic polarization $Q_m=U_m=V_m=0$ and intensities  $I_m$ varying from $1$ to $20$ flux units. The positions of the sources are shown in Fig. \ref{skymod} while the circles sizes indicate the flux ratio between the apparent and true flux values. The number $M=50$ was chosen to emulate a typical situation with a LOFAR observation at about $150$ MHz with an average beam diameter of about $8$ degrees. At much higher frequencies, the beams are narrower and the sources are less bright, therefore, 'clustering' \citep{Kaz1} of sources may be required to get sufficient directions along which to calibrate.

\begin{figure}[htbp]
\begin{minipage}{1.00\linewidth}
\centering
 \centerline{\epsfig{figure=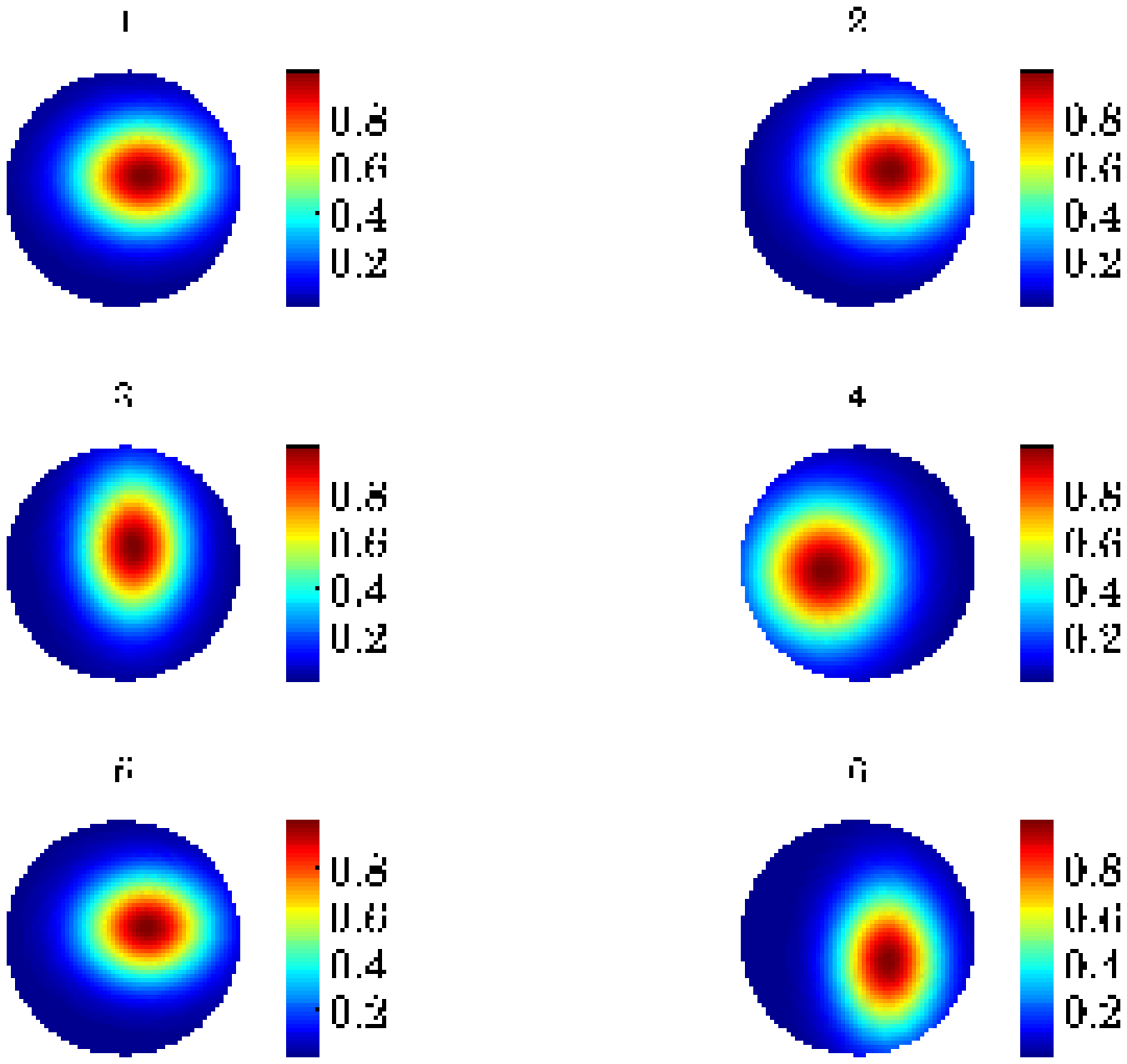,width=8.0cm}}
\vspace{0.1cm}\centerline{(a)}\smallskip
\centering
 \centerline{\epsfig{figure=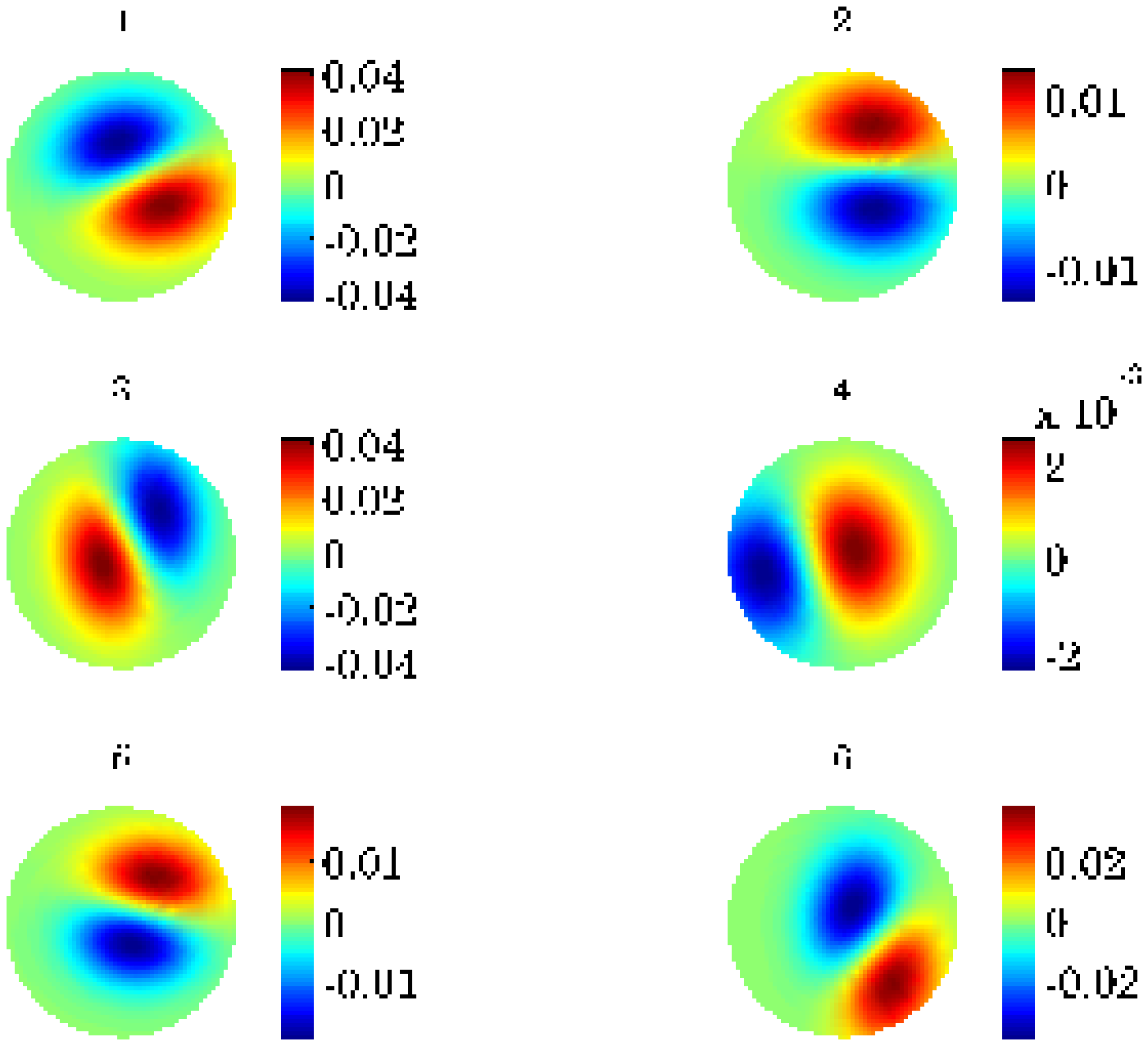,width=8.0cm}}
\vspace{0.1cm}\centerline{(b)}\smallskip
\caption{Original beams: (a) Real value (b) Imaginary value. Each beam is a Gaussian in amplitude with a randomly chosen major and minor axis and a random offset from the center. The amplitude is multiplied by a random linear phase screen to make the beam complex.\label{orig_beams}}
\end{minipage}
\end{figure}

\begin{figure}[htbp]
\begin{minipage}{1.00\linewidth}
\centering
 \centerline{\epsfig{figure=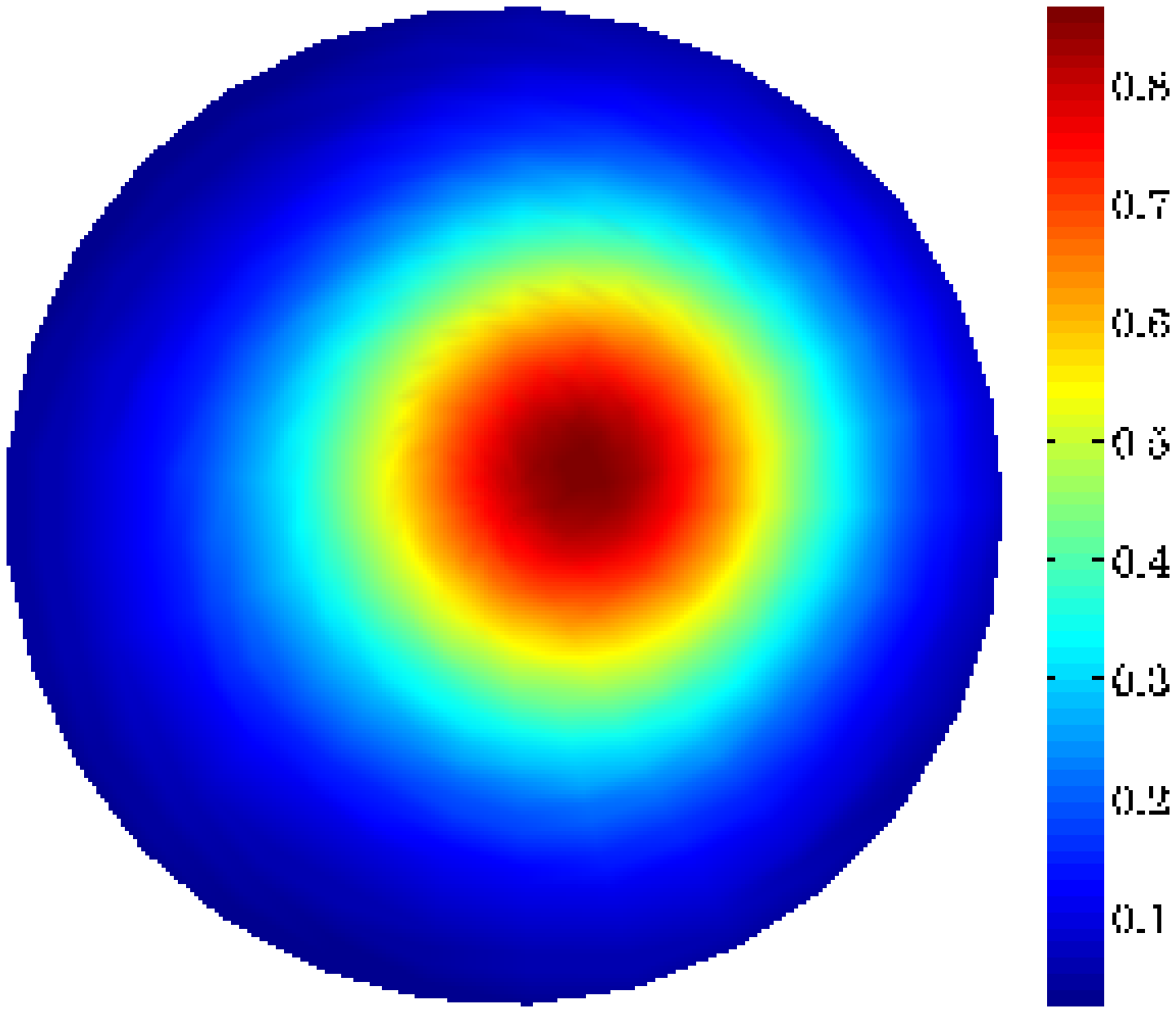,width=8.7cm}}
\caption{Beam shape used  for calculating the apparent sky fluxes. We start with a real beam so the imaginary value is zero.\label{initial_beam}}
\end{minipage}
\end{figure}

We simulate an interferometer with $N=6$ stations. The beam shape of each station is generated to be a Gaussian with random major and minor axes and random offsets from the center of the field of view, as shown in Fig. \ref{orig_beams}. In addition, we multiply this with a random linear phase screen to make the beam complex. In order to generate the apparent sky model, we attenuate the intrinsic fluxes of the sky model with the mean of the amplitude of the beam shapes shown in Fig. \ref{orig_beams}, as this is what the fluxes that will be seen in an image made by this interferometer. The mean beam shape is shown in Fig. \ref{initial_beam}. We also corrupt the apparent fluxes with Gaussian noise, having zero mean and a variance of $0.01$.

\begin{figure}[htbp]
\begin{minipage}{1.00\linewidth}
\centering
 \centerline{\epsfig{figure=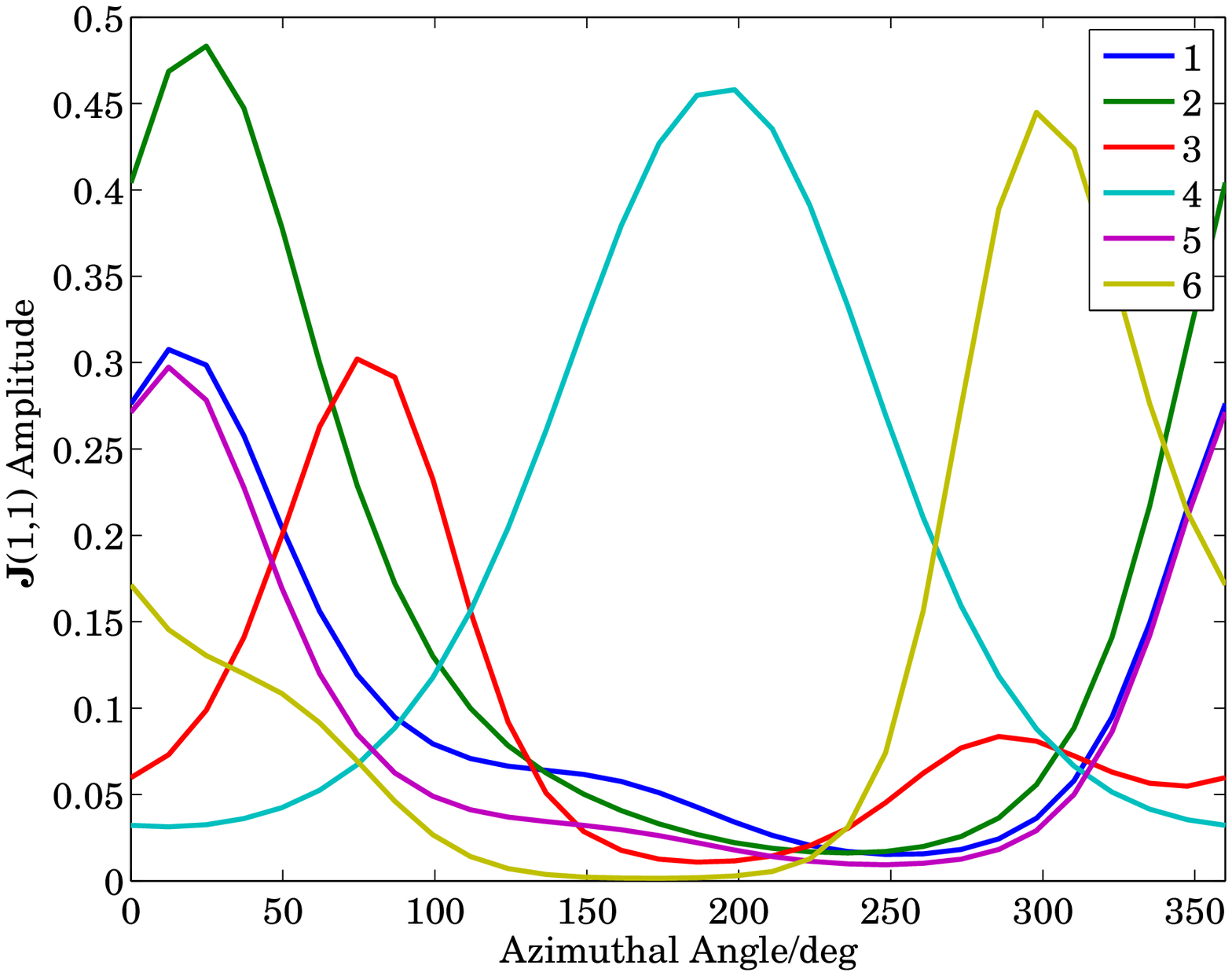,width=8.0cm}}
\vspace{0.1cm}\centerline{(a)}\smallskip
\centering
 \centerline{\epsfig{figure=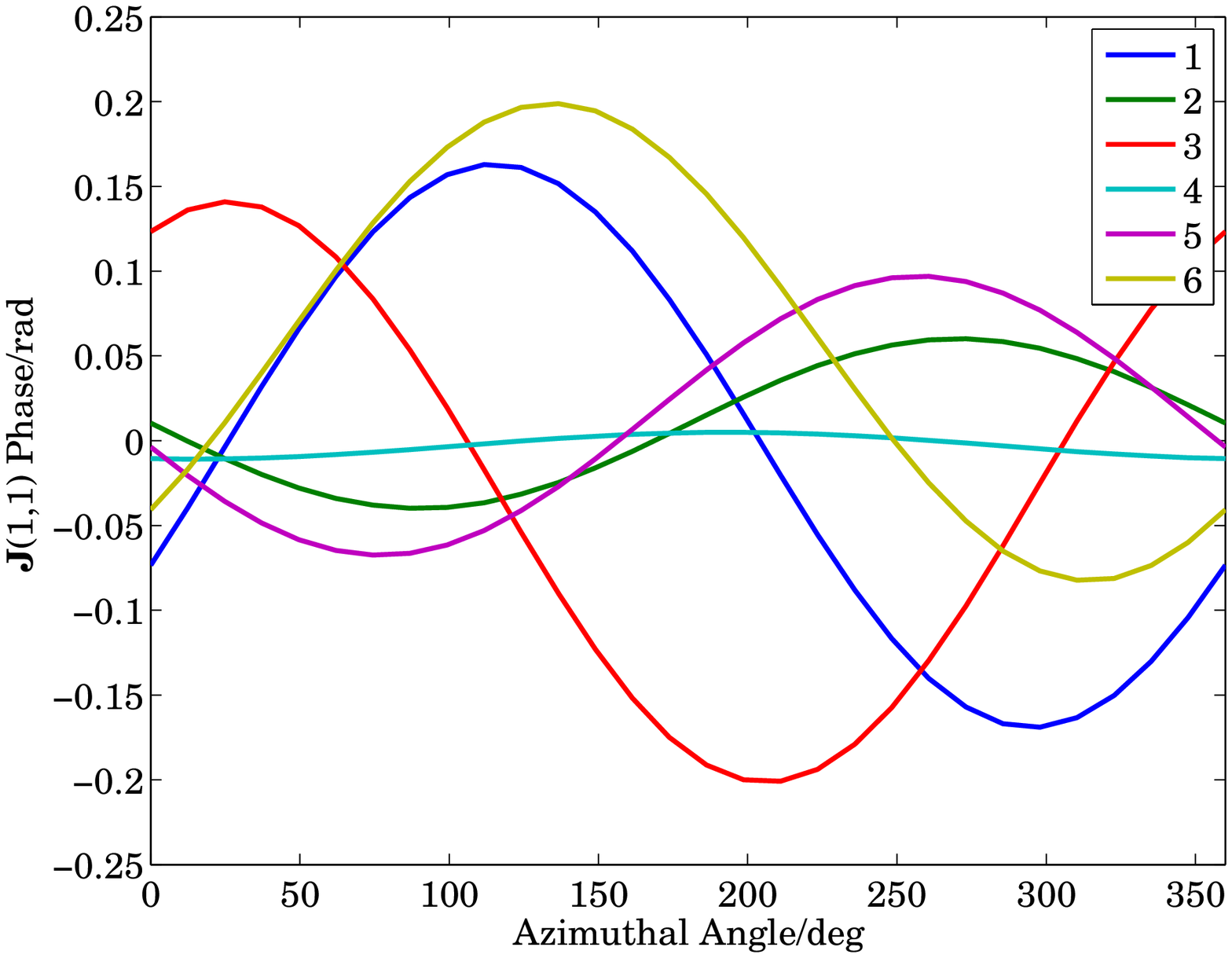,width=8.0cm}}
\vspace{0.1cm}\centerline{(b)}\smallskip
\caption{Variation of the beam gain along an azimuthal track: (a) Amplitude (b) Phase. The beam gain is calculated for all $6$ stations along a direction $4$ degrees away from the center of the field of view.\label{beams_gains}}
\end{minipage}
\end{figure}

Once we have generated the apparent sky model, we calculate the gain along each direction using the true beam shape and the apparent flux. As an example, we give the gain variation along an azimuthal track for a direction $4$ degrees away (in zenith angle) from the field center in Fig. \ref{beams_gains}. We only show the $(1,1)$ entry of the matrix ${\bf J}_{pm}$ in Fig. \ref{beams_gains}. For each direction $m$, we calculate the true Jones matrix ${\bf J}_{pm}$ and use a randomly generated unitary matrix ${\bf U}_m$ to get the values used in (\ref{fcost}) as ${\bf J}_{pm}{\bf U}_m+{\bf N}$, where ${\bf N}$ is a complex Gaussian noise matrix, with elements having zero mean and a variance of $0.01$. For a real observation, this step is replaced by calibration along the direction of each source in the sky model.

We consider minimizing (\ref{fcost}) by the proposed method as well as by the unconstrained Broyden Fletcher Goldfarb Shanno (BFGS) optimization routine \citep{NW}. For both routines, we need to supply an initial value for ${\bf B}$. We consider the initial beam for all stations to be a circular Gaussian with major and minor axes diameter of $4$ degrees (half the field of view). We selected spherical harmonics with order $4$ as the basis functions for ${\bf B}$. Therefore, there are $D=16$ basis functions and the size of ${\bf B}$ is $6\times 16$. The initial value of ${\bf B}$ was used to calculate the value for $\alpha=trace({\bf B}^H{\bf B})$. Spherical polar coordinates, centered at the pole of Fig. \ref{skymod} are used to calculate the basis functions.

\begin{figure}[htbp]
\begin{minipage}{1.00\linewidth}
\centering
 \centerline{\epsfig{figure=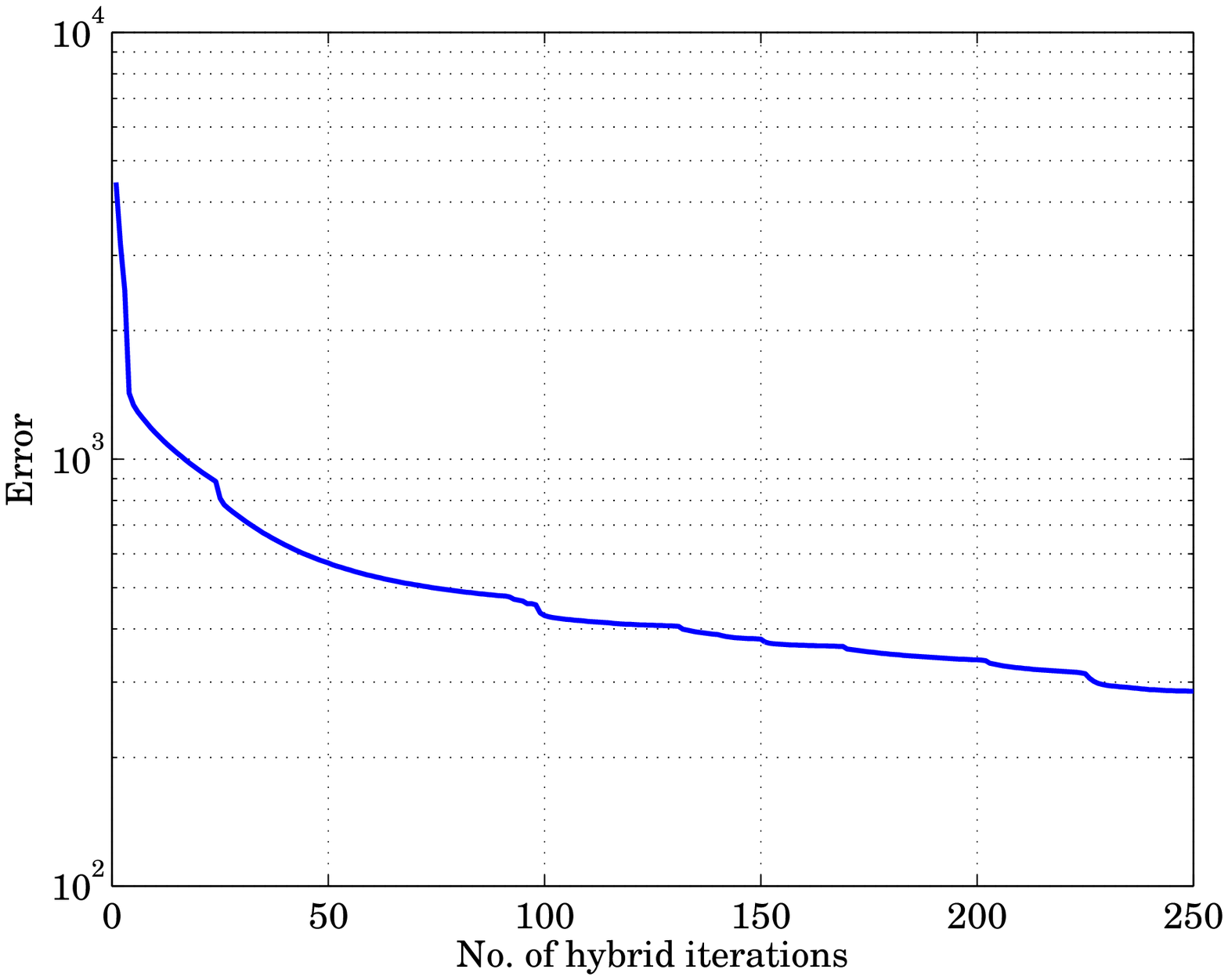,width=8.0cm}}
\vspace{0.1cm}\centerline{(a)}\smallskip
\centering
 \centerline{\epsfig{figure=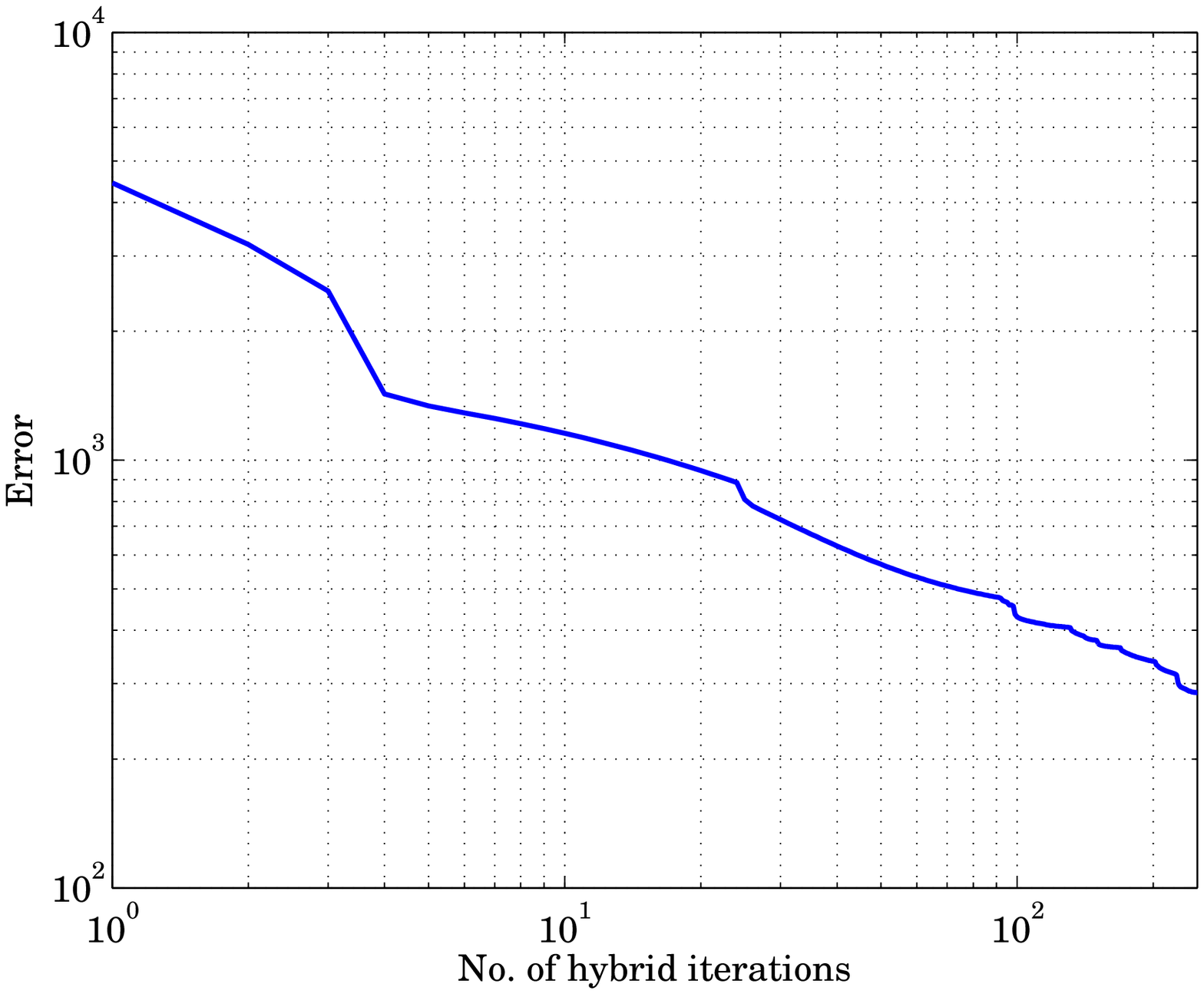,width=8.0cm}}
\vspace{0.1cm}\centerline{(b)}\smallskip
\caption{Reduction of the cost function with the number of iterations for the proposed algorithm. (a) linear-log scale (b) log-log scale. The sudden jumps of the cost occur when RBFGS finds a successful solution.\label{cost_iter}}
\end{minipage}
\end{figure}

The reduction of the cost function with the number of hybrid iterations of the proposed algorithm is shown in Fig. \ref{cost_iter}. At certain points of the iteration, RBFGS algorithm finds successful solutions and the cost is reduced at a rate which is superlinear.

\begin{figure}[htbp]
\begin{minipage}{1.00\linewidth}
\centering
 \centerline{\epsfig{figure=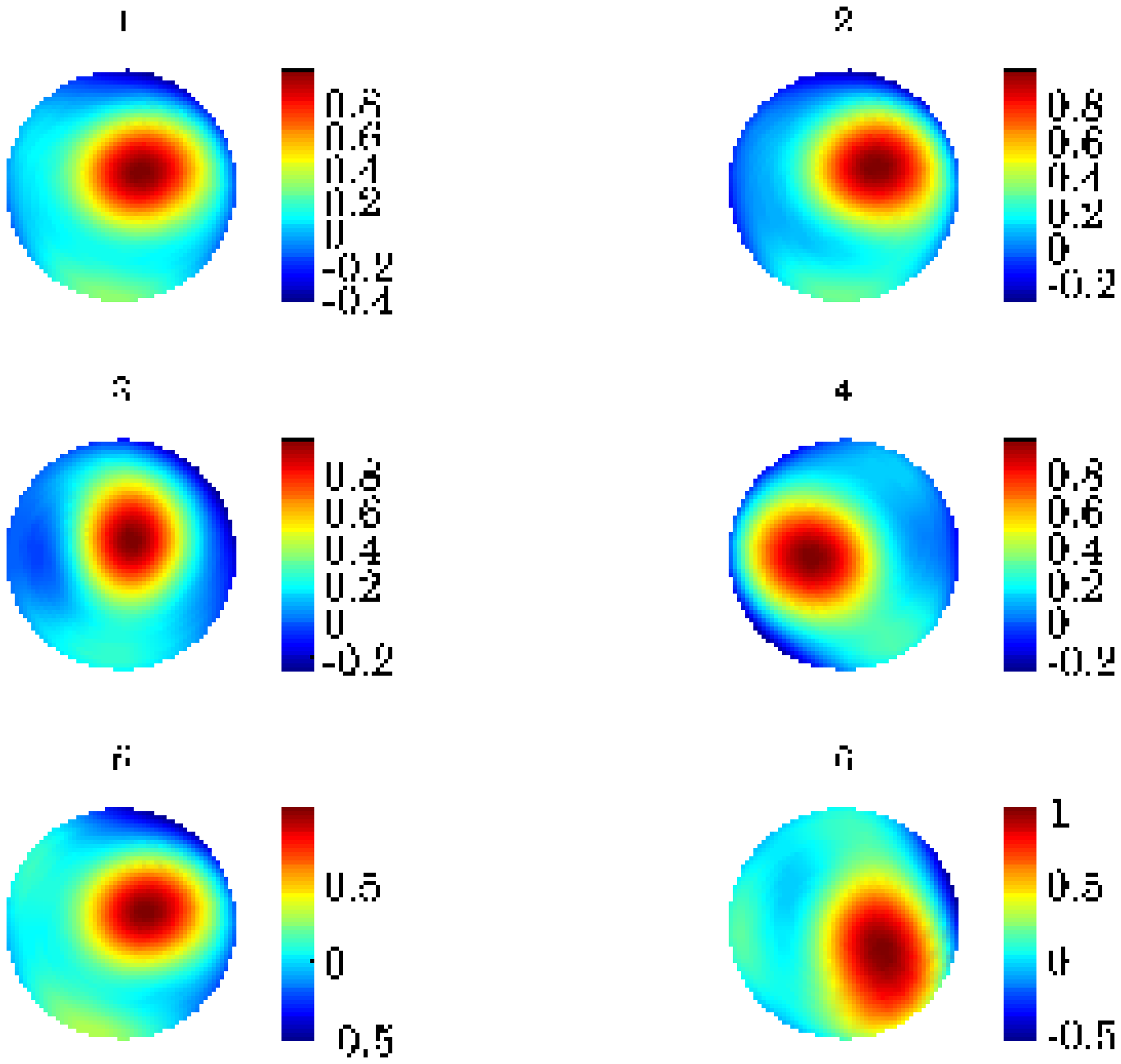,width=8.0cm}}
\vspace{0.1cm}\centerline{(a)}\smallskip
\centering
 \centerline{\epsfig{figure=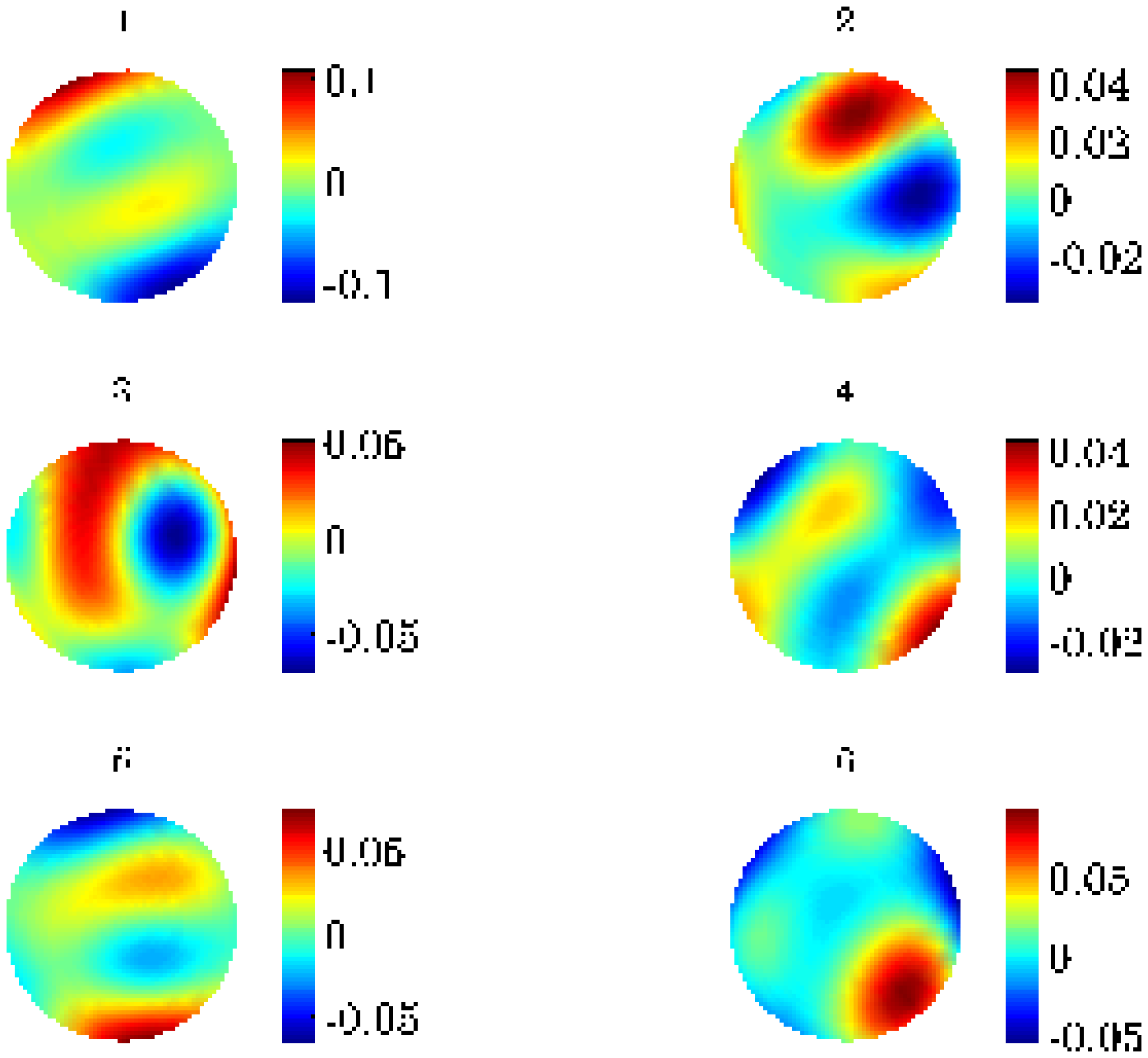,width=8.0cm}}
\vspace{0.1cm}\centerline{(b)}\smallskip
\caption{Estimated beams using unconstrained BFGS optimization: (a) Real value (b) Imaginary value.  The total squared error between the original beams and the estimated ones is about $152$.\label{unconstrained_beams}}
\end{minipage}
\end{figure}

\begin{figure}[htbp]
\begin{minipage}{1.00\linewidth}
\centering
 \centerline{\epsfig{figure=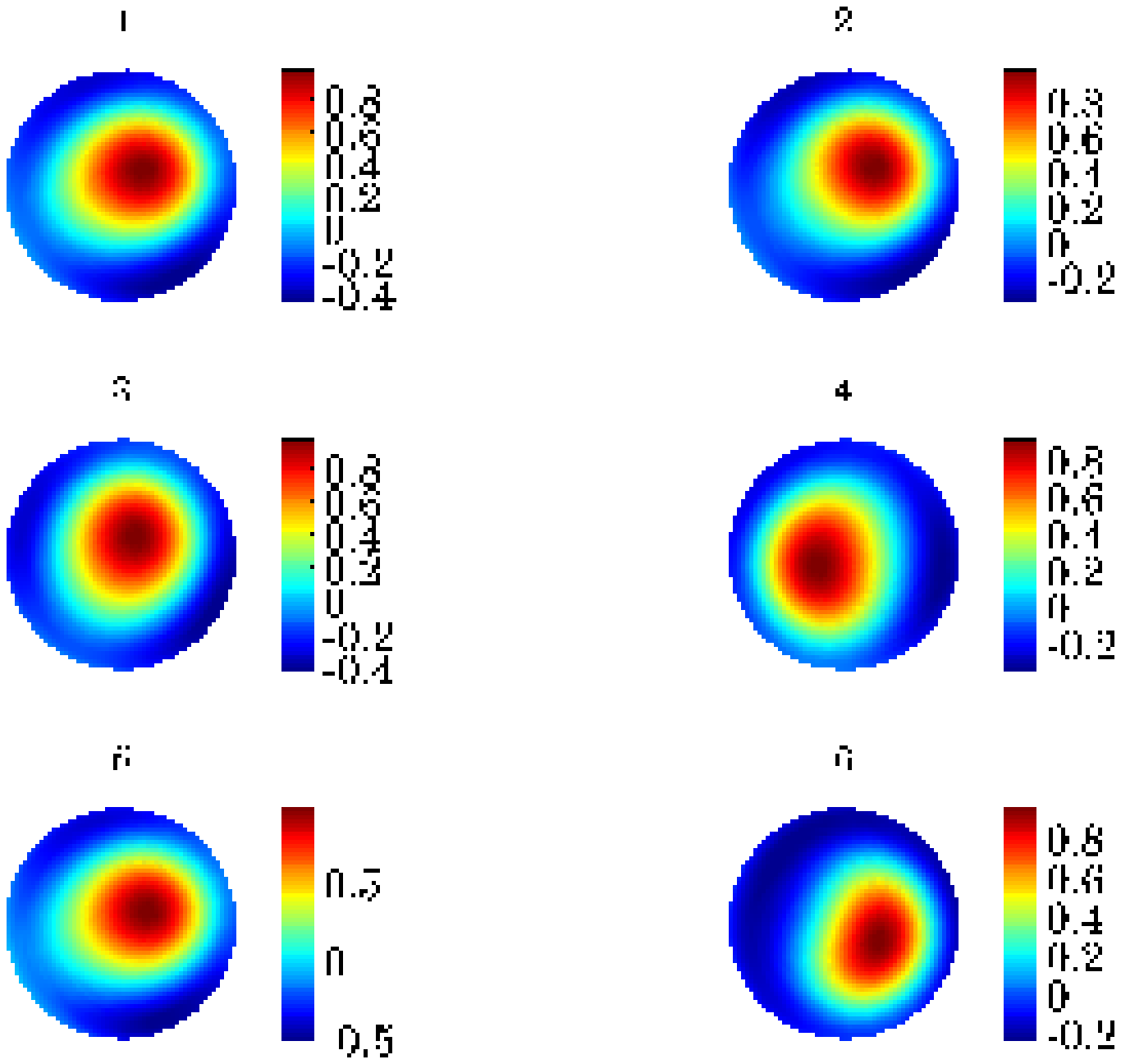,width=8.0cm}}
\vspace{0.1cm}\centerline{(a)}\smallskip
\centering
 \centerline{\epsfig{figure=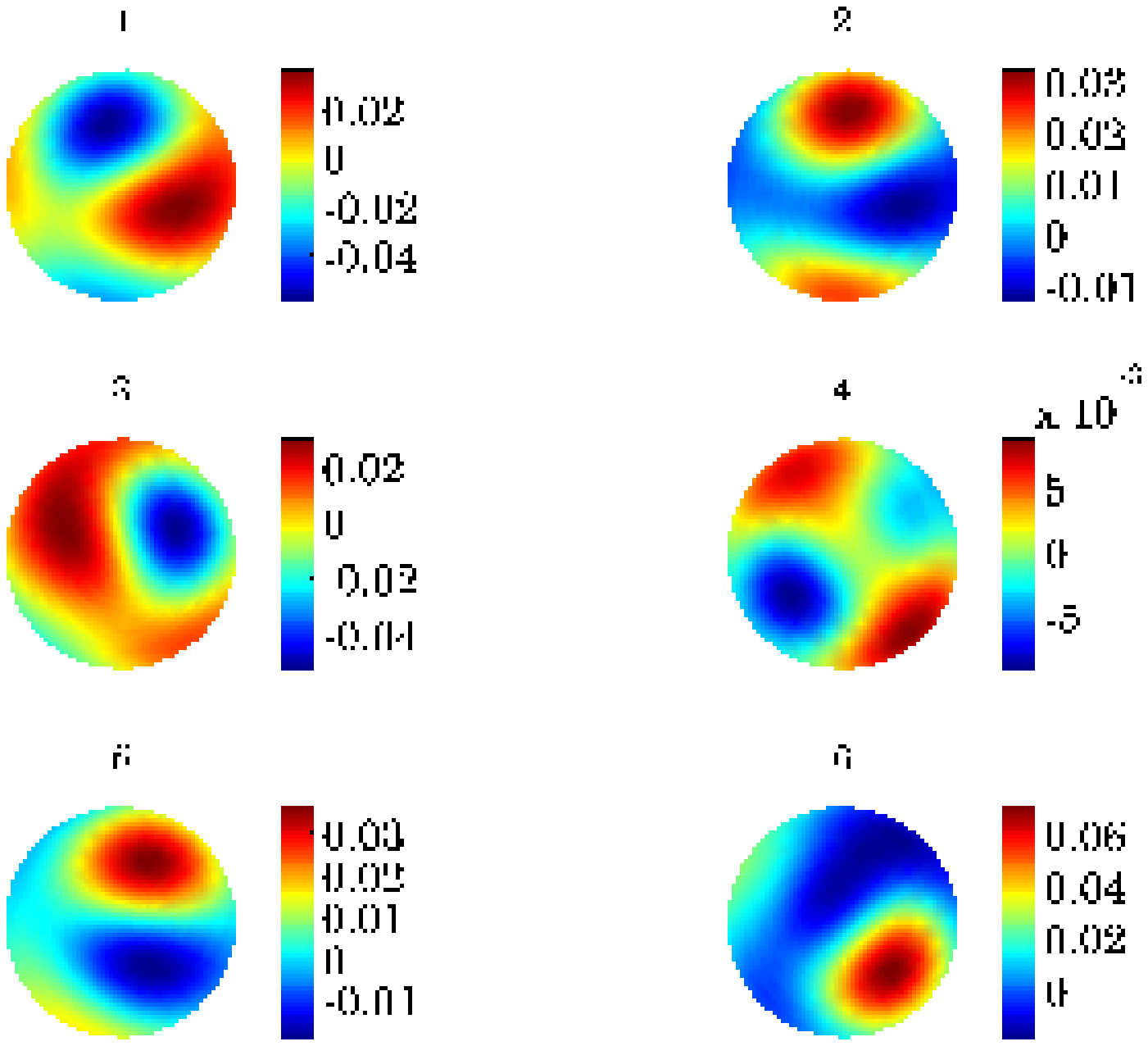,width=8.0cm}}
\vspace{0.1cm}\centerline{(b)}\smallskip
\caption{Estimated beams using the proposed algorithm. (a) Real value (b) Imaginary value. The total squared error between the original beams and the estimated ones is about $130$.\label{constrained_beams}}
\end{minipage}
\end{figure}

In Fig. \ref{unconstrained_beams}, we have given the results of the unconstrained optimization with $2500$ iterations. The results of the proposed algorithm is shown in Fig. \ref{constrained_beams}, after $250$ hybrid iterations. The inner iterations used is $10$ so the total number of iterations for the proposed method is $2500$ as well.

\begin{figure}[htbp]
\begin{minipage}{1.00\linewidth}
\centering
 \centerline{\epsfig{figure=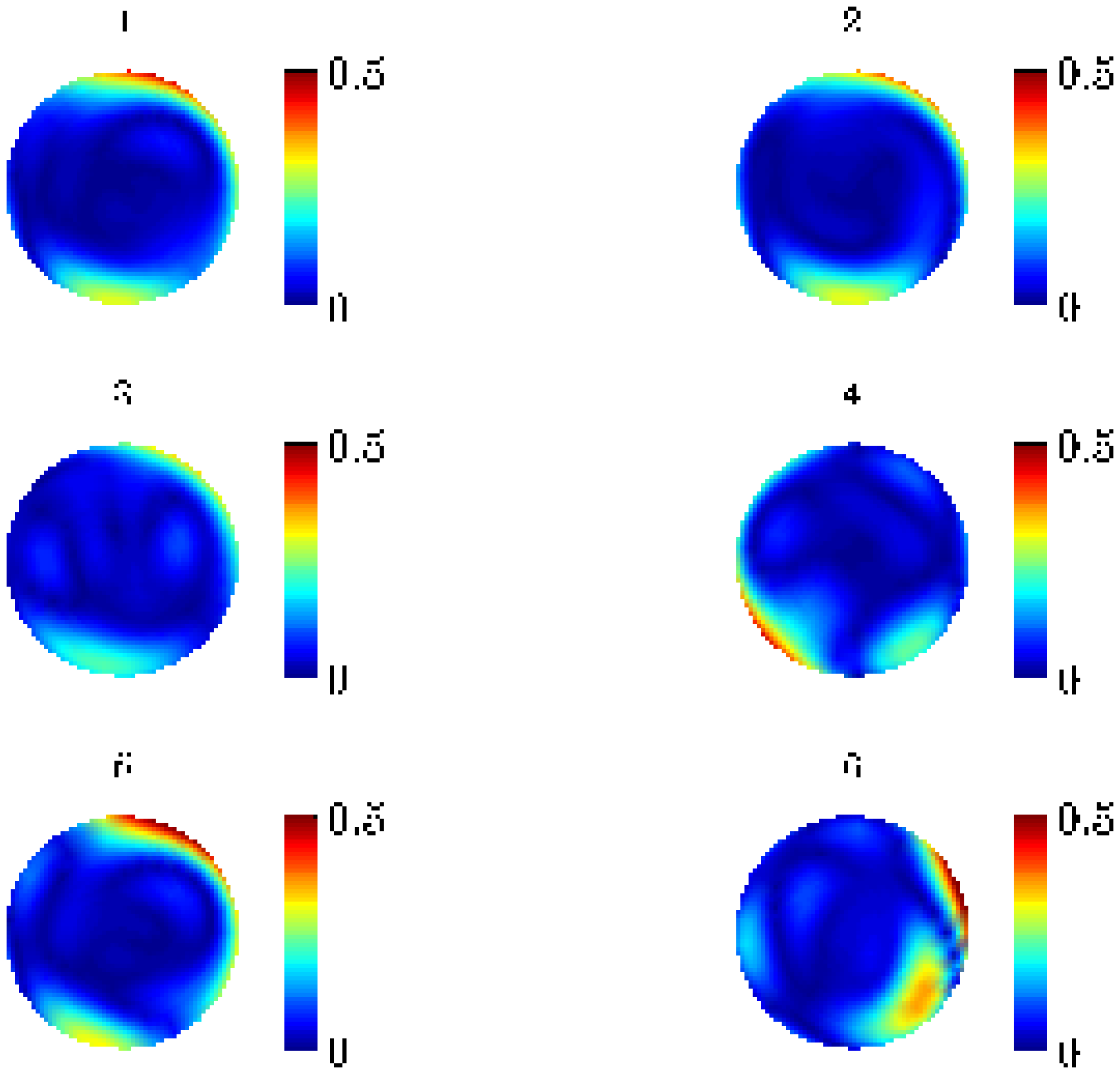,width=8.0cm}}
\vspace{0.1cm}\centerline{(a)}\smallskip
\centering
 \centerline{\epsfig{figure=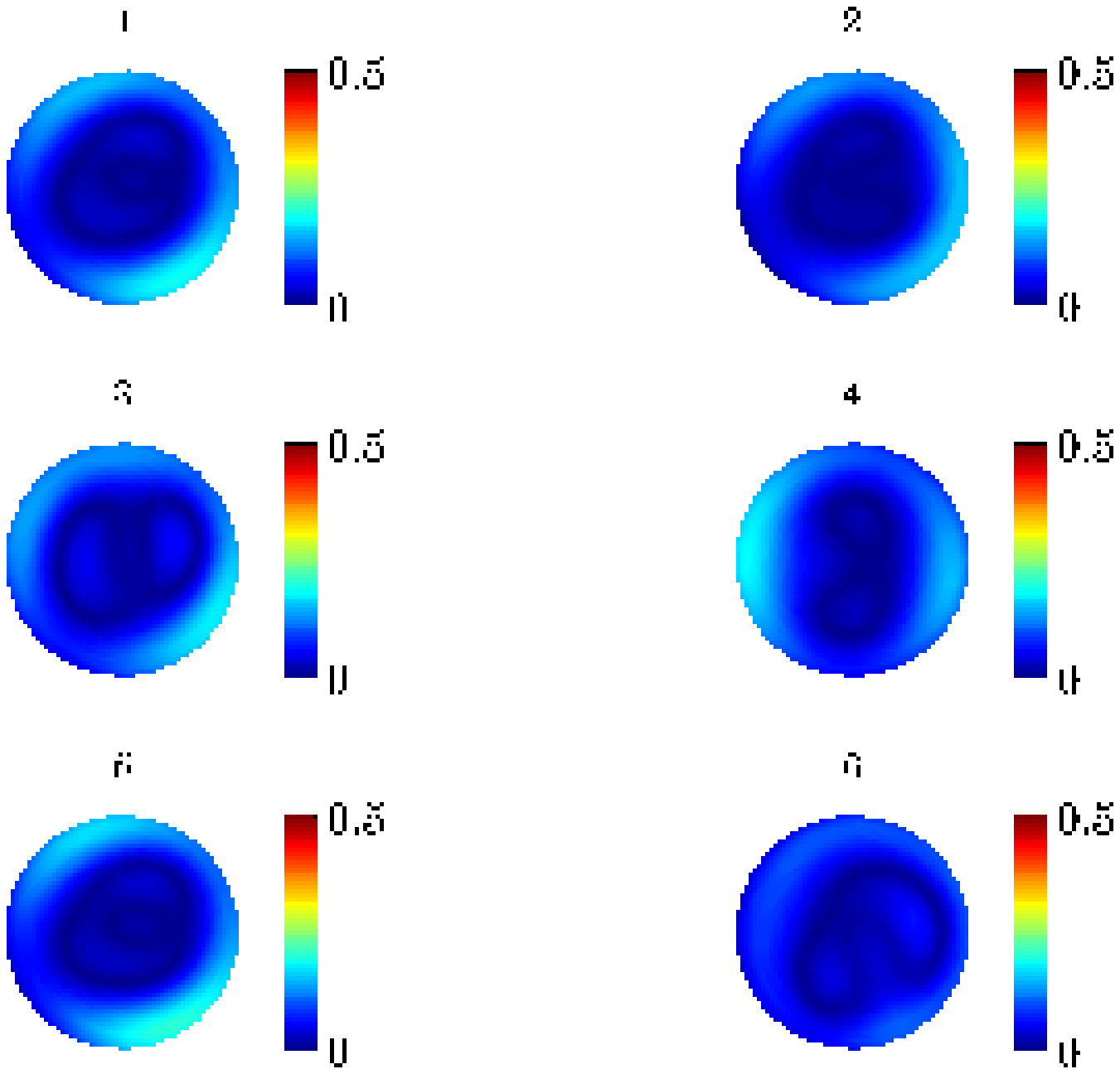,width=8.0cm}}
\vspace{0.1cm}\centerline{(b)}\smallskip
\caption{Error amplitude between estimated and original beams. (a) Unconstrained estimate (b) Proposed estimate. There are fewer outliers in the error of the proposed approach.\label{beam_err}}
\end{minipage}
\end{figure}

Comparison of Figs. \ref{unconstrained_beams} and \ref{constrained_beams} with the original in Fig. \ref{orig_beams} clearly shows the superiority of the proposed method. The real values of the beams in Figs. \ref{unconstrained_beams} and \ref{constrained_beams} indicate that the proposed method gives a more focused beam shape as opposed to the unconstrained approach. Moreover, the proposed method recovers the imaginary value of the beams better than the unconstrained approach. Note that the beam number $4$ in Fig.  \ref{orig_beams} has almost zero imaginary value (implying that the phase component is negligible). While the proposed approach also gives a very small value for this beam in Fig. \ref{constrained_beams}, the unconstrained approach gives a significantly higher value, as seen in Fig. \ref{unconstrained_beams}. In Fig. \ref{beam_err}, we also show the error amplitude between estimated beams and the original beams. For a quantitative comparison, we have calculated the total squared error between the original beams and estimated beams for the full field of view, sampled at $30\times30$ grid points. For the unconstrained case, we get a total error of $152$ while for the proposed case we only get an error of $130$. Comparison of the final cost of $f({\bf B})$ in (\ref{fcost}) at the end of each algorithm shows a different result. With conventional optimization, we get a much lower cost for (\ref{fcost}) compared with the proposed method. This is clearly a misleading result due to the ill-posedness of the problem.

\begin{figure}[htbp]
\begin{minipage}{0.98\linewidth}
\centering
 \centerline{\epsfig{figure=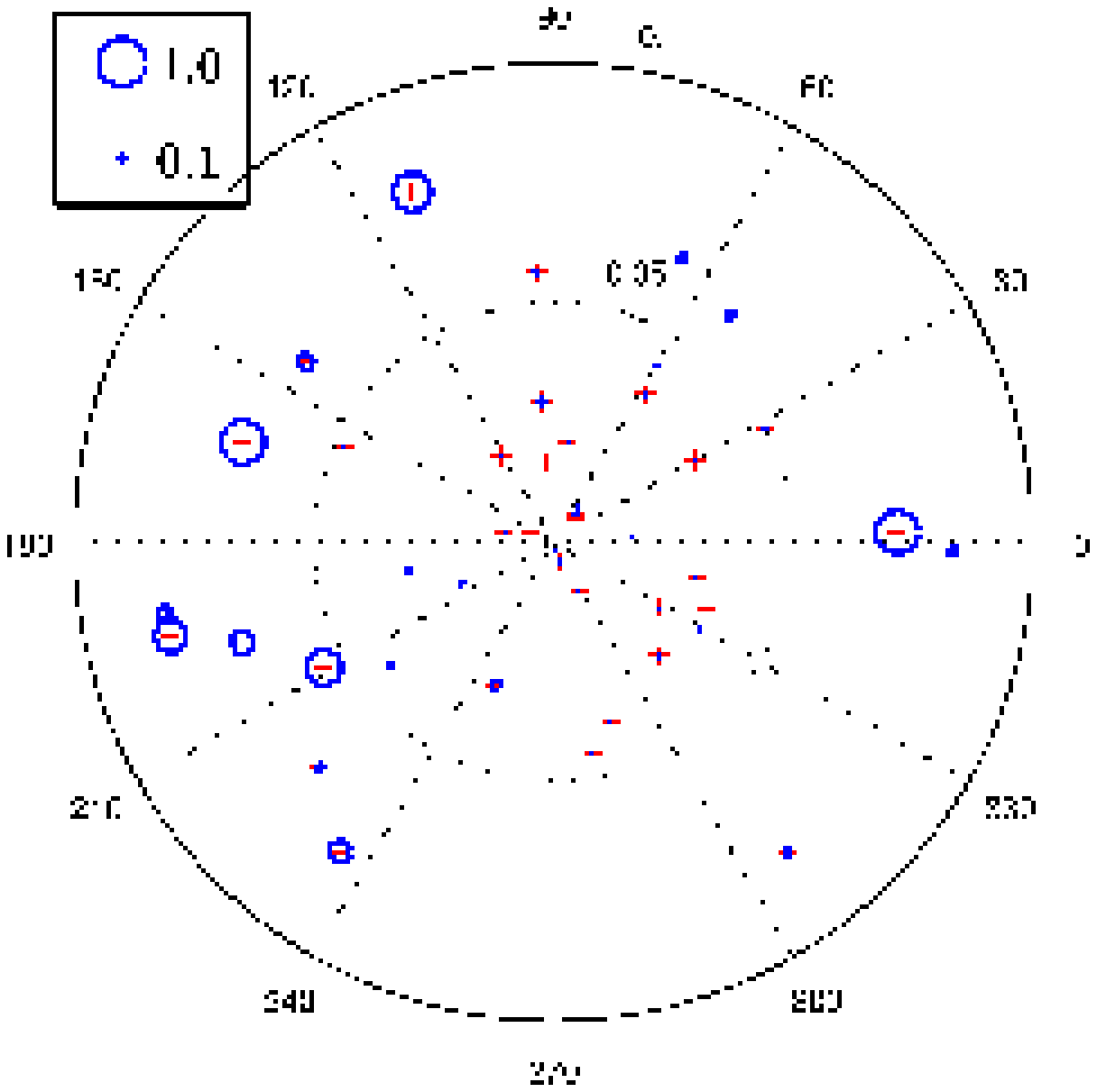,width=10.0cm}}
\caption{The error (difference between estimated flux and true flux) in estimated fluxes using the beam shape obtained by unconstrained optimization. The size of the circles represent the magnitude of the absolute error.\label{flux_err_uncons}}
\end{minipage}
\end{figure}

\begin{figure}[htbp]
\begin{minipage}{0.98\linewidth}
\centering
 \centerline{\epsfig{figure=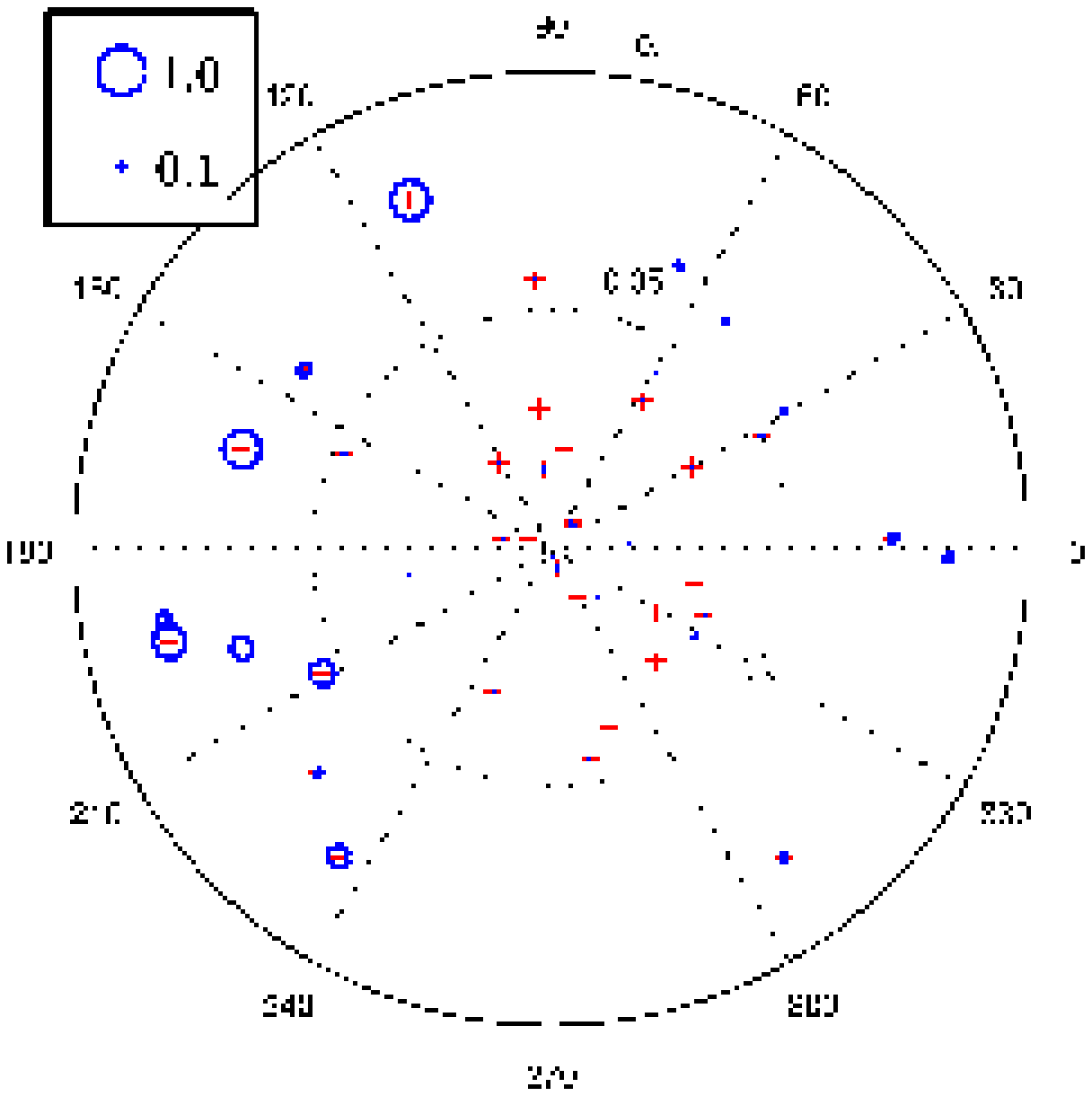,width=10.0cm}}
\caption{The error (difference between estimated flux and true flux) in estimated fluxes using the beam shape obtained by the proposed approach. The size of the circles represent the magnitude of the absolute error.\label{flux_err_cons}}
\end{minipage}
\end{figure}

In Figs. \ref{flux_err_uncons} and \ref{flux_err_cons}, we have shown the error in estimating the intrinsic flux using (\ref{cohest}). Both figures show the difference of the estimated flux with the true flux by the size of the circles. In Fig. \ref{flux_err_uncons}, we have used the beam estimate obtained using the unconstrained approach while in Fig. \ref{flux_err_cons} we have used the beam shape obtained using the proposed approach. Both beam estimates give good recovery of the true fluxes within the inner region of the field of view.

The sources at the outlier clearly shows an error in the recovered flux, mainly because their apparent flux is low and are more susceptible to noise. Therefore, in order to improve the flux estimation of outlier sources, we can use diversity in frequency and time. Because the sky is almost invariant, we can combine beam estimates obtained over different time and frequency intervals to improve the flux estimates of outlier sources.

\section{Conclusions\label{conc}}
We have proposed a method of estimating interferometer beam shapes used in a radio interferometric observation by using the directional gains obtained towards known celestial sources. This ill posed problem is solved using optimization on a Riemannian manifold. As compared with conventional optimization, the proposed method give better results. However, the proposed method is computationally more expensive than conventional (unconstrained) optimization. Future work will address the application of this method to real interferometric observations and reducing the computational cost.

\section{Acknowledgments}
We thank the anonymous reviewers for the careful review and helpful comments that enabled us to enhance this paper.

\bibliographystyle{spbasic}
%\bibliography{manref}

\begin{appendix}
\section{Proof of (\ref{alpha}): \label{proof_alpha}}
Let the total power received by all stations be $\Gamma$. We can express this as\beq
\Gamma=\sum_{p,m} | \gamma_{pm} |^2 = \sum_{p,m} |{\bf e}_p^T {\bf B} {\bf b}_m|^2
\eeq
where $p$ is summation over all stations $(1,\ldots,N)$ and $m$ is summation over an infinite number of directions in the sky that covers the full field of view. Note that the summation over $m$ is not restricted to the directions where we have known sources.
We have 
\beq
\Gamma=trace\left( {\bf B} \sum_m ( {\bf b}_m {\bf b}_m^H {\bf B}^H \sum_p {\bf e}_p{\bf e}_p^T) \right)
\eeq
and using the fact that
\beq
\sum_p {\bf e}_p {\bf e}_p^T ={\bf I}
\eeq
we get
\beq
\Gamma=trace\left( \sum_m ( {\bf b}_m {\bf b}_m^H ) {\bf B}^H {\bf B}  \right).
\eeq
Let
\beq
 {\bmath \Upsilon}{\bmath \Upsilon}^H=\sum_m ( {\bf b}_m {\bf b}_m^H ).
\eeq
Then,
\beq
\Gamma=trace\left( ({\bf B} {\bmath \Upsilon})^H ({\bf B}  {\bmath \Upsilon})\right) = || {\bf B}  {\bmath \Upsilon} ||^2 \le || {\bf B} ||^2  ||{\bmath \Upsilon} ||^2 
\eeq
Taking into account that $||{\bmath \Upsilon} ||^2$ is fixed for a given basis, we can keep $\Gamma$ below a certain level by keeping
\beq
|| {\bf B} ||^2=trace\left( {\bf B}^H {\bf B} \right) = \alpha
\eeq
where $\alpha$ is a fixed real value. One additional point to be raised here is that by selecting an orthonormal basis, we get ${\bmath \Upsilon}\approx {\bf I}$, therefore an orthonormal basis is always preferred (although in practice hard to realize).

\section{Proof of (\ref{gradf}): \label{proof_gradf}}
We can rewrite (\ref{fcost}) as 
\beq
f({\bf B})=\sum_{p,q,m} trace\left(({\bf X}_{pqm}\otimes{\bf Z}_{pqm}-{\bf Y}_{pqm}\otimes {\bf 1})^H({\bf X}_{pqm}\otimes{\bf Z}_{pqm}-{\bf Y}_{pqm}\otimes {\bf 1})\right)
\eeq
where ${\bf X}_{pqm}\buildrel\triangle\over={\bf C}_{pqm}$, ${\bf Z}_{pqm}\buildrel\triangle\over={\bf e}_{p}^T{\bf B}{\bf b}_m {\bf b}_m^H{\bf B}^H{\bf e}_q$, ${\bf Y}_{pqm}\buildrel\triangle\over={\bf J}_{pm} \widetilde{\bf C}_{pqm}{\bf J}_{qm}^H$ and ${\bf 1}\buildrel\triangle\over=1$.
This can be simplified as
\beqn
f({\bf B})=&&\sum_{p,q,m} trace({\bf X}_{pqm}^H{\bf X}_{pqm})trace({\bf Z}_{pqm}^H{\bf Z}_{pqm})\\\nonumber
&\mbox{}& -trace({\bf X}_{pqm}^H{\bf Y}_{pqm})trace({\bf Z}_{pqm}^H)\\\nonumber
&\mbox{}& -trace({\bf Y}_{pqm}^H{\bf X}_{pqm})trace({\bf Z}_{pqm})\\\nonumber
&\mbox{}&+trace({\bf Y}_{pqm}^H{\bf Y}_{pqm}).
\eeqn
Using \cite{Mdiff}, we can take the derivative of each term with a trace of ${\bf Z}_{pqm}$ to yield (\ref{gradf}).

\end{appendix}
\end{document}